\documentstyle[12pt]{ioplppt} 
\begin{document}
\title{Closed-form solutions of the Schr\"odinger equation 
for a class of smoothed Coulomb potentials}

\author{Charles W. Clark} 
\address{Electron and Optical Physics Division, National 
Institute of Standards and Technology, Technology 
Administration, U. S. Department of Commerce,
Gaithersburg, MD 20899} 

\begin{abstract}
An infinite family of closed-form solutions
is exhibited for the 
Schr\"odinger equation for the potential 
$V({\bf r}) = -Z/\sqrt{|{\bf r}|^{2} + a^{2}}$.  
Evidence is presented for an approximate dynamical
symmetry for large values of the angular momentum $l$.

\end{abstract}
\jl{2}
\submitted
\maketitle

\section{Introduction}
\label{sec:intro}

A recent paper \cite{LC92} demonstrated the existence of a 
family of closed-form solutions to the one-dimensional 
Schr\"odinger equation for the potential 

\begin{equation}
V_{1}(x) = -\frac{1}{\sqrt{x^2 + a^2}}
\label{1Dpot}
\end{equation}
which is widely used \cite{JES88,CC90,BKPR91,GLR91,SE91,EC96,LC96} 
in the modeling of atomic response to strong time-dependent 
radiation fields.   The approach used in Ref. \cite{LC92} was 
somewhat analogous to the Sturmian method \cite{MR70} for 
generating a complete basis set of Coulomb wavefunctions,
in which the energy $E$ is held fixed and the
nuclear charge $Z$ is determined as an eigenvalue.  
Ref. \cite{LC92} determined ``eigenvalues'' $a_n$ of the cutoff 
parameter $a$, for which the eigenenergies $E_n$ of
the Schr\"odinger equation for the potential Eq. (\ref{1Dpot})
take the values 
 
\begin{equation}
E_n = -\frac {1}{2n^2}; 
\label{hydrogenenergies}
\end{equation}

\noindent these are the energies of the states of the three-
dimensional
hydrogen atom, in the usual system of atomic units (used 
throughout the present paper) in which 
the numerical value of unity is assigned to the 
mass $m$ and charge $e$ of the electron
and to the reduced Planck's constant $\hbar$.
The associated eigenfunctions $\Psi_n(x)$ were found
to have a closed-form expression:
\begin{equation}
\Psi_n(x) = x^{\nu} e^{-\kappa\chi} f_n(\chi) ,
\label{1Deq}
\end{equation}
where: $\nu$ = 0 or 1 for cases of even and odd parity 
respectively; $\chi = \sqrt{x^2 + {a_n}^2}$; 
$f_n$ is a polynomial of degree $n \ge 1 - \nu$; 
$\kappa = 1/n$; and ${a_n}^2$ is a root of an $n^{th}$-degree
polynomial.

In the present work a similar approach is applied to the 
three-dimensional smoothed Coulomb potential

\begin{equation}
V({\bf r}) = -\frac{Z}{\sqrt{|{\bf r}|^{2}+a^{2}}}.
\label{3Dpot}
\end{equation}
This potential arises in the Kramers-Hennenberger 
transformation 
of the equations
of motion of a hydrogen atom in a radiation field 
\cite{KH,DF}, 
and it
is qualitatively similar to some pseudopotentials used in
density-functional calculations of electronic structure
(e.g. Ref. \cite{BHS}).  It is shown here that, 
subject to an assumption similar to one made in 
Ref. \cite{LC92}, the Schr\"odinger equation for this 
potential admits an infinite number of closed-form solutions 
for each value of the angular momentum $l$.  
The lowest-energy solutions for each $l$ 
are quite simple, so knowledge of these exact 
results may be useful for 
calibrating numerical methods that must 
be used to solve general equations of this type.
To the best of the author's knowledge, these are
the first exact results for Schr\'odinger
eigenvalues and eigenfunctions for potenials of the
class described by Eq. (\ref{3Dpot}).

\section{Method of solution}
\label{sec:method}

The extension of the method of \cite{LC92} is  
straightforward. 
We proceed from the Schr\"odinger equation

\begin{equation}
 -\frac{1}{2} \nabla^2 \Psi({\bf r}) - \frac{Z}{\sqrt{|{\bf 
r}|^{2}+a^{2}}} \Psi({\bf r}) = E \Psi({\bf r}),
\label{BasicSch}
\end{equation}
and invoke the usual spherical coordinate factorization 
$\Psi({\bf r}) =
\psi_l(r) Y_{lm}(\theta,\phi)$.  With the definitions $\alpha 
= Za,
\rho = \sqrt{(Zr)^2 + \alpha^2}, \epsilon = E/Z^2 = 
-\kappa^2/2$, and

\begin{equation}
\psi_l(r) = r^l \phi_l(\rho),
\label{transformation}
\end{equation}
the equation

\begin{equation}
\left(1 - 
\frac{\alpha^2}{\rho^2}\right)\frac{\partial^2\phi_l(\rho)}
{\partial \rho^2} + \left(\frac{\alpha^2}{\rho^3} + \frac{2l 
+ 
2}{\rho}\right)
\frac{\partial\phi_l(\rho)}{\partial\rho} +
\left(\frac{2}{\rho} - \kappa^2\right)\phi_l(\rho) = 0. 
\label{basiceqn}
\end{equation}
is obtained.  We now postulate a solution of the form

\begin{equation}
\phi_l(\rho) = e^{-\kappa\rho}\sum_{i = 0}^{n}c_i\rho^i ,
\label{seriesExpansion}
\end{equation}
where $n$ is an integer to be determined.
For economy of expression the dependence of $\kappa$ and
$c_i$ upon $n$ and $l$ is not made explicit 
in the notation, but it is held to be implicit.

We now show that Eq. (\ref{seriesExpansion})
does not provide a general solution,  but that it is applicable 
for discrete values of $\alpha$.  To see this, we substitute 
eq. (\ref{seriesExpansion}) in the l.h.s. of Eq. 
(\ref{basiceqn}), and require the net coefficient
of each power of $\rho$ to vanish identically.  The 
coefficient of 
$\rho^j$ that emerges from this operation is

\begin{eqnarray}
-\alpha^2(j+4)(j+2)c_{j+4} + 
\alpha^2\kappa(2j+5)c_{j+3} + \nonumber \\
\left((j+2)(j+2l+3)-\alpha^2\kappa^2\right)c_{j+2} + 
2\left(1-\kappa(j+l+2)\right)c_{j+1} = 0
\label{coefficientvanish}
\end{eqnarray}
for $j = -3, -2, -1, . . ., n-1$.  Thus we must solve the 
$n+3$ 
simultaneous equations (\ref{coefficientvanish}) in the $n+2$ 
variables
$\{c_1, c_2, . . ., c_{n}, \kappa, \alpha\}$.  

We proceed by treating Eq. (\ref{coefficientvanish}) as a 
{\em four-term} recurrence relation, and solve for $c_{j}$
sequentially downwards from $j = n-1$. Because $c_{j} = 0$
for $j > n$, the solution for $j = n - 1$ gives

\begin{equation}
\kappa = \frac{1}{n+l+1}. 
\label{kappaeqn}
\end{equation}
Thus $\kappa$ is uniquely determined by $n$ and $l$.
Eq. (\ref{kappaeqn}) describes
exactly the spectrum of the three-dimensional Coulomb potential,
i.e. the limiting case of Eq. (\ref{BasicSch}) with $a = 0$;
in that case $n$ coresponds to the 
number of nodes in the radial eigenfunction.
We shall see subsequently that, for $\alpha \ne 0$,
we must have $n \geq 1$, and the maximum number of radial nodes 
in the wavefunction
described by Eqs. (\ref{seriesExpansion}) and (\ref{kappaeqn})
is $n - 1$.  The potential of Eq. (\ref{3Dpot}) 
has a long-range
Coulomb tail and a non-Coulombic component at small $r$,
so its Schr\"odinger spectrum is naturally described
in the language of quantum defect theory
\cite{Seaton}.  In that terminology, Eq. (\ref{kappaeqn})
describes a state with an integer value of 
the quantum defect $\mu$, which is necessarily negative
for $a \ne 0$. The following development will indicate that
all eigenfunctions with integral quantum defects obtained
with potentials of the class Eq. (\ref{3Dpot}) are described
by Eq. (\ref{seriesExpansion}).

Since Eq. (\ref{basiceqn}) is homogeneous, we can set
$c_{n} = 1$ without loss of generality. The values of
$c_i$ for $i < n$ are then determined in terms of $c_{n}$ by 
downward recursion using Eq. (\ref{coefficientvanish}). 
For $j = n-2$, we get

\begin{equation}
c_{n-1} = \frac {1}{2} \left( n(n+l+1)(n+2l+1) - 
\frac{\alpha^2}{n+l+1}\right) c_{n} = 
p_1(\alpha^2)c_{n}, 
\label{firstpolynomial}
\end{equation}
where $p_1(x)$ designates a first-degree polynomial in $x$.
From this equation it is apparent that there will be a 
solution
for $n = 0$ only if $\alpha = 0$, which is the familiar 
Coulombic case.

Inspection of the structure of Eq. (\ref{coefficientvanish}) 
shows
that by continuing the recursion process downward in
$j$ we get 
  
\begin{equation}
c_{n-m} = p_m(\alpha^2)c_{n}, 
\label{mpolynomial}
\end{equation}
with $p_m(\alpha^2)$ being a polynomial of degree $m$ in
$\alpha^2$.  So for $j = 0$ and $j = -1$ we find

\begin{equation}
c_1 = p_{n-1}(\alpha^2)c_{n} \qquad {\rm and} \qquad  c_0 = 
p_{n}(\alpha^2)c_{n} \\  
\label{01polynomial}
\end{equation}
respectively.  The last two cases to be considered are 
$j = -2$ and $j = -3$.  These are found to give
the {\em same} equation:
\begin{equation}
c_1 = \kappa c_0 = \frac {c_0}{n+l+1}.
\label{negative}
\end{equation}
Thus, from Eqs. (\ref{01polynomial}) and 
(\ref{negative}), we see that Eq. (\ref{seriesExpansion}) will
provide a valid solution if
\begin{equation}
q_{n}(\alpha^2) = p_{n}(\alpha^2) - (n+l+1) 
p_{n-1}(\alpha^2) = 0
\label{determinesalpha}
\end{equation}
or, in other words, if $\alpha^2$ is a root of the $n^{th}$ 
degree
polynomial $q_{n}$.

The applicability of Eq. (\ref{seriesExpansion}) thus depends
upon some roots of Eq. (\ref{determinesalpha}) being positive 
real numbers.  The investigation reported here has not 
uncovered a general proof that Eq. (\ref{determinesalpha})
has any such roots, but calculations
carried out for $l \leq 10^6$ and $n \leq 20$
suggest that {\em all} its roots are 
positive real numbers and are nondegenerate.  Let us adopt
this as a hypothesis.  If it is true, then 
the following statements hold:
 \\
 \\
\noindent {\bf i.}  {\em For each $l, n$ there are $n+1$ 
values of       
$\alpha^2$ for which Eq. (\ref{BasicSch}) has solutions of the
form Eqs. (\ref{transformation}, \ref{seriesExpansion})}.  This
includes the previously-known (Coulombic) value 
$\alpha^2 = 0$, plus the $n$ roots of the polynomial $q_{n}$.
 \\
 \\
\noindent {\bf ii.}  {\em The only potentials of the class 
eq. (\ref{3Dpot}) which have bound states with Coulombic energies 
are just those with values of $\alpha^2$ that are solutions
to Eq. (\ref{determinesalpha}).}  This is because Eq. (\ref{3Dpot})
describes a monotonic function of $\alpha^2$.  Thus, for a 
given $l$, its 
associated discrete Schr\"odinger eigenvalues will increase 
uniformly towards zero as $\alpha$ increases.  
For a given $l$, $Z$, and Coulombic energy, 
\begin{equation}
E_{n l} = -\frac{1}{2} \frac{Z^2}{\left(n+l+1\right)^2},
\label{hydrogenic}
\end{equation}
there will
be some maximum value of $\alpha^2$ for which $E_{n l}$ 
occurs as an eigenvalue, i.e., that for which it is the lowest 
eigenvalue.  As $\alpha^2$ is decreased from this maximum, 
$E_{n l}$ will next occur in the spectrum when it is the 
second lowest eigenvalue, then
as the third lowest, etc., until finally at 
$\alpha^2 = 0$,  when it is the ($n+1$)-th lowest eigenvalue.  
Thus there are indeed only
$n+1$ values of $\alpha^2$ for which Eq. (\ref{hydrogenic}) occurs
in the spectrum, which is consistent with the stated 
hypothesis, and if the hypothesis is true, these values 
must thus coincide with the roots of 
Eq. (\ref{determinesalpha}) 

The argument ii. is illustrative of the actual results of
computations of solutions of Eqs. 
(\ref{coefficientvanish}) and (\ref{determinesalpha}),
as will be described below.  The largest value of $\alpha^2$ 
for a given ($n,l$) is associated with a nodeless 
eigenfunction; the next largest value of $\alpha^2$, 
with an eigenfunction with one node; and so on to the smallest 
nonzero value of $\alpha^2$, which corresponds to an 
eigenfunction with $n-1$ nodes.

\section{Results}
\label{sec:Results}

The eigenvalues of $\alpha^2$ can be easily be found for a 
given ($n,l$) by numerical solution of the polynomial 
equation (\ref{determinesalpha}).  They have relatively
simple closed forms for $n = 1$ and 2, which are presented
here.  Numerical tables are given below for $1 \le l \le 3$ and
$n \le 10$.

\subsection{$n = 1$}
\label{sec:nequal1}
For $n = 1$ we obtain
\begin{equation}
 \alpha^2 = 2(l+2)^3, \qquad c_{0} = l + 2, \qquad c_{1} = 1, 
\label{alpha1}
\end{equation}
so that the (nonnormalized) solution of Eq. (\ref{BasicSch}) is
\begin{equation}
\psi_l(r) = r^l e^{-\rho/(l+2)}\left(l + 2 + \rho\right),
\label{psi1}
\end{equation}
with $\rho = \sqrt{(Zr)^2 + 2 (l + 2)^3}$, and 
$a^2 = Z^{-2}\alpha^2 = Z^{-2}2(l+2)^3$ .

\subsection{$n = 2$}
\label{sec:nequal2}
For $n = 2$ there are two solutions for $\alpha^2$:
\begin{equation}
\alpha^2 = \left[3 \pm \sqrt{\frac{l+15}{l+3}}\right](l+3)^3 .
\label{alpha2}
\end{equation}
With the choice of $c_2 = 1$, we can write the
expressions for the two sets of coefficients $c_i$ in the
common form:
\begin{equation}
c_0 = \frac{\alpha^2}{2} - (l + 3)^2 (2 l + 3), \hfill 
c_1 = \frac{c_0}{l + 3} , \hfill
\label{cvals1}
\end{equation}
with the value of $\alpha^2$ to be chosen as appropriate.

\subsection{$n > 2$}
\label{sec:ngt2}

Although closed-form expressions can be obtained for
$\alpha^2$ and $c_i$ for $n = 3$ and 4, they have the
typical cumbersome form of roots of cubic and quartic
equations, and it does not seem particularly useful to 
record them here in full.  However, a simplifying relationship is
worth noting.  For these values of $n$, the values 
of $\alpha^2$ can be written as
\begin{equation}
\alpha^2 = (n + l + 1)^3 (n+ 1 + \beta),
\label{definesbeta}
\end{equation}
\noindent where $\beta$ is a root of the polynomials,
\begin{equation}
\beta^3 -4 \frac{l+19}{l+4} \beta - 24 \frac{l+14}{(l + 
4)^2} = 0
\label{beta3}
\end{equation}
\begin{equation}
\beta^4 - 10 \frac{l+23}{l+5} \beta^2 - 48 
\frac{3l + 50}{(l+5)^2} \beta + 9 \frac{(l^3 + 51 l^2 + 643 l + 945)}
{(l+5)^3} = 0 ,
\label{beta4}
\end{equation}
\noindent for $n=$ 3 and 4, respectively.
In the limit of large $l$, the solutions of Eqs. 
(\ref{beta3}) and (\ref{beta4}) tend respectively to
$\beta = 0, \pm 2$ 
and $\beta = \pm 1, \pm 3$.  Thus from Eqs.
(\ref{alpha1}-\ref{beta4}) we see that 
for large $l$, the smallest value of $\alpha^2$ tends to
$\alpha^2 = 2(n+l+1)^3$ for $n = 1$ through 4.  This motivates
the choice of Eq. (\ref{definesbeta}) as a general
representation for the values of $\alpha^2$,  and it has
been used to record those values in the tables given below.
It has been found that, to a high degree of numerical accuracy,
the computed values of $\beta$ for a given ($n,l$) sum
to zero, so that substituting $\beta = 0$ in 
eq. \ref{definesbeta} apparently locates 
$\alpha^2 = (n + l + 1)^3 (n+ 1)$ as the 
average of the values of $\alpha^2$.  No fundamental explanation
of this apparent fact is advanced here.

Tables \ref{table1}, \ref{table2}, \ref{table3} give the 
values of $\beta$ for 
$n \leq 10$ for $l = 1 -3$.  A similar set of values for
$l = 0$ can be found in Table 1 of Ref. \cite{LC92}, so they
are not repeated here \cite{Comment}.

\subsection{Systematic behavior of wavefunctions}
\label{sec:systematics}

Numerical calculations indicate that the wavefunctions
described by Eq. (\ref{seriesExpansion}) exhibit the qualitative behavior discussed at the 
end of Sec. \ref{sec:method}.  For a given $(n,l)$, denote by
$\alpha_k$ the $k^{th}$ smallest value of $\alpha$ obtained
in solving Eq. (\ref{determinesalpha}), with $k = 1, 2, . . ., n$.
Numerical experiments indicate that the wavefunction corresponding
to $\alpha_k$ has $n_r = n-k$ radial nodes, a pattern that 
was observed in the one-dimensional cases treated in 
ref. \cite{LC92}.  An example of this behavior is illustrated 
in Fig. \ref{wavefunctions} for the case
$n = 10$, $l = 1$.

Fig. \ref{alfavals} depicts the values of $\alpha_k$ for 
$0 \leq l \leq 4$ as a function of $n_r+l+2 = n^*$, the
effective principal quantum number of atomic spectroscopy,
which is related to the energy via Eq. (\ref{kappaeqn}).  For a 
given ($l,k$), as labelled in Fig. \ref{alfavals}, a discernable
sequence of values of $\alpha_k$ is observed; these sequences
are seen to approach definite limits as $n*$ increases.
This is related to the well-known phenomenon in
atomic spectroscopy in which quantum defects
tend to constant values 
high in Rydberg series.  The slow variations of 
high-$n$ quantum defects are due to the presence of
a fixed, short-range, non-Coulombic part of the 
potential experienced by a Rydberg electron.
Correspondingly, the fixed value of quantum defect
obtained in the present method is associated with
slow variation of $\alpha_k$ at large $n$. 

For large $\alpha^2$ and large $k$, on the other hand,
the eigenfunctions approach those of the three-dimensional
harmonic oscillator.  This can be seen by expanding
$Z/\sqrt{r^2+a^2}$ in powers of $r/a$ for large $a$; 
retaining the lowest two terms gives the Schr\"odinger
equation for the harmonic oscillator.  If first-order
perturbation theory is used to include the effects
of the $r^4$ term in this expansion, we obtain
the approximate spectrum,
\begin{eqnarray}
E_{n_r l} \rightarrow - \frac {Z}{a} + 
\left[ 2n_r+l+3/2 \right] \sqrt{\frac {Z}{a^3}} \nonumber \\
- \frac {3}{8 a^2} \left[6 n_r \left(n_r+l+3/2 \right) + 
\left(l+5/2 \right) \left(l+3/2 \right)\right] , 
\end{eqnarray}
as $a \rightarrow \infty$.
Fig. \ref{correlationdiagram} 
is a correlation diagram that
displays the connection between this limit and the
hydrogenic limit $a = 0$, if one keeps the number 
of radial nodes, $n_r$, fixed as $a$ varies. Two familiar 
cases of $l$-degeneracy are apparent in this
figure: $n_\rho + l$ = constant for the hydrogen atom,
and $2n_\rho + l$ = constant for the three-dimensional
harmonic oscillator.

\subsection{The spectrum for large values of $l$ and $a$}
\label{sec:largel}

In sec. \ref{sec:ngt2} it was mentioned that as
$l \rightarrow \infty$, we find $\alpha_n \rightarrow 2 (n+l+1)^3$
for $1 \leq n \leq 4$.  Thus in this limit we
recover a case of near-$l$-degeneracy
similar to that encountered in hydrogen: 
there are values of $\alpha$ that support
degenerate eigenfunctions with different
values of $l$, described as a class by the
equation $n_r +l =$ constant.  
The approach to this limit is relatively slow, 
apparently like $l^{-1}$
as suggested by Eq. (\ref{alpha2}): e.g. for $l = 10,000$ and 
$n \leq 10$, the actual value of $\alpha$ changes by about a
part in $10^4$ for a unit change in $n_r$ at constant $n_r+l$.
Numerical experiments suggest that this approximate
degeneracy is a general phenomenon at large $l$.

There is a simple effect of this kind of degeneracy
for all potentials that have a long-range Coulomb tail 
and some non-Coulombic behavior localized 
at small $r$.  In such systems, the centrifugal 
barrier presented to a high-$l$ 
wavefunction - commonly called a
``nonpenetrating orbital'' - will prevent it from 
sampling the non-Coulombic region.  Thus the Coulombic 
$l$-degeneracy is largely undisturbed 
for large $l$.  The quantum defects $\mu_l$
of nonpenetrating orbitals tend to
zero as $l \rightarrow \infty$, a phenomenon
that is universal in actual atomic systems, where,
apart from isolated instances of series perturbation,
observed quantum defects are hardly greater than 0.01
for $l \geq 5$ .

However, the effect 
encountered in the present system is quite different.
The non-Coulombic behavior of the potential extends to
very large $r$, so the large-$l$ eigenfunctions
are substantially modified from their Coulombic forms:
their quantum defects are negative integers.
The appearance of this novel $l$-degeneracy
presumably derives from the existence of a constant
of motion for the Schr\"odinger equation 
given by Eq. (\ref{BasicSch}) that emerges in the large-$l$ limit,
but it has not
been identified in the present work.  

\section{Conclusions}
\label{sec:conclusions}

A simple algebraic method has been presented to generate
an infinite number of parameters $a$ for which closed-form
solutions may be found to the Schr\"odinger equation
for the class of smoothed Coulomb potentials described
by Eq. (\ref{3Dpot}).  The procedure
bears some superficial resemblances to the Sturmian approach.
However, it is not based on a system of
orthogonal polynomials, and because the functions it
generates are obviously incomplete, it probably cannot
be simply related to known orthogonal systems.
These results should be useful for testing, to arbitrary
numerical accuracy, methods that integrate
the Schr\"odinger equation for Coulomb-like systems,
such as are encountered in electronic structure and
collision problems.  The approach also points to the
possibility of previously unknown integrals of
motion in these systems.

%

\Figures
\begin{figure} 
\caption{
Solutions for $n = 10$, $l = 1$: 
wavefunctions $\psi_1(\rho)$ (solid line) and 
scaled potentials 
$Z^2 V(r) = -\left[ \rho^2 + \alpha^2 \right]^{-\frac {1}{2}}$ (dashed line) 
vs. $\rho$ (horizontal axis). The numerical value of $\alpha$ is displayed for each case. The horizontal range $0 \le \rho \le 350$ 
is the same for all figures, and the negative portion of
the vertical axis that is displayed covers the range [-0.11,0]
in all cases.  The wavefunctions $\psi_1(\rho)$ are not normalized, 
and have each been scaled to fit the frame; they have also been chosen
to be positive near $\rho = 0$, a convention that differs 
trivially from
that used in the constructive procedure presented in
sec. \protect{\ref{sec:method}}, where instead the 
long-range tail is taken to be positive.
}
\label{wavefunctions} 
\end{figure}

\begin{figure}
\caption{
Plot of $\alpha$ vs. $n^* = n_r+l+2$ 
for $0 \leq l \leq 4$,  
$2 \leq n+l+1 \leq 16$, and $\alpha \leq 100$.  Key: plus: $l = 0$;
x: $l = 1$; star: $l = 2$; square: $l = 3$; triangle: $l = 4$.  The first 
few series ($l$,$k$) are explicitly labelled.}
\label{alfavals} 
\end{figure}
 
\begin{figure} 
\caption{
Correlation diagram for the spectrum of 
eq. (\protect{\ref{BasicSch}}) as a function of $a$
(increasing schematically to the right),
reflecting the conservation of $n_r$.  The
vertical position of a level is proportional
to its effective principal quantum number $n$*,
which determines the energy via $E = -Z^2/(2n^{*2})$
or $E = -Z/a + (n^{*}+3/2) \protect{\sqrt{Z/a^3}}$ for the hydrogenic
and oscillator limits, respectively; the horizontal
position corresponds to $l$ as indicated.  The values
of $n$* for the two limits are displaced and on
different scales to ease visualization of the reordering
of levels. 
}
\label{correlationdiagram} 
\end{figure}
%
%

\begin{table}
\caption{
Values of $\beta$ as defined in 
Eq. (\protect{\ref{definesbeta}}), for $l = 1$,
$3 \leq n \leq 10$.  
The $n$ values of $\beta$ for each 
$n$ are given in ascending order in the
columns as labeled.
}
\label{table1}
\begin{tabular}{rrrr}
\br
$n = 3$&4&5&6\\
\mr
-3.436527918374&-4.654288835724&-5.773876395778&-6.844547165256\\
-0.954320351864&-3.021850153582&-4.650636370300&-6.042758624724\\
4.390848270238&0.696058817445&-1.972562395354&-4.058664349356\\
&6.980080171861&2.712910374971&-0.480918834922\\
&&9.684164786461&4.966113875958\\
&&&12.460775098300\\
\end{tabular}
\begin{tabular}{rrrr}
\br
$n = 7$&8&9&10\\
\mr
-7.888799456947&-8.917837293622&-9.937641169524&-10.951590892028\\
-7.298711510457&-8.472195004982&-9.593574030219&-10.680835341450\\
-5.793704911519&-7.307257906333&-8.675819112578&-9.946495139814\\
-3.007931174911&-5.101818236107&-6.904370845863&-8.505293514974\\
1.322982663950&-1.607950021737&-4.050370418227&-6.148266454200\\
7.380032760853&3.353214980339&0.059395253271&-2.710246349993\\
15.286131629031&9.908007968597&5.551945760159&1.934250402020\\
&18.145835513845&12.519780081038&7.879187144780\\
&&21.030654481943&15.194895308886\\
&&&23.934394836774\\
\end{tabular}
\end{table}

\begin{table}
\caption{Values of $\beta$ as defined in
Eq. (\protect{\ref{definesbeta}}), for $l = 2$,
$3 \leq n \leq 10$ ,
presented as in Table \protect{\ref{table1}}
}
\label{table2}
\begin{tabular}{rrrr}
\br
$n = 3$&4&5&6\\
\mr
-3.278125698868&-4.521965213374&-5.668426333247&-6.761249358836\\
-0.798234614319&-2.785523299847&-4.411763596407&-5.826275313135\\
4.076360313187&0.779586310517&-1.735936565934&-3.774065245823\\
&6.527902202704&2.700152211531&-0.287039774645\\
&&9.115974284057&4.853558868056\\
&&&11.795070824383\\
\end{tabular}
\begin{tabular}{rrrr}
\br
$n = 7$&8&9&10\\
\mr
-7.822716061200&-8.864931236359&-9.894830804857&-10.916571653805\\
-7.110457925768&-8.311162452760&-9.456572802320&-10.564284465555\\
-5.507083989124&-7.038000908935&-8.431024341412&-9.727562574162\\
-2.723146982242&-4.785217058704&-6.586194666454&-8.200598472288\\
1.453378478484&-1.351453753382&-3.733658908457&-5.808927517424\\
7.172075276697&3.411667214518&0.271140940945&-2.413663893356\\
14.537951203153&9.611814936391&5.536500893854&2.092437435636\\
&17.327283259231&12.143135917875&7.791260127102\\
&&20.151503770825&14.745312392743\\
&&&23.002598621108\\
\end{tabular}
\end{table}

\begin{table}
\caption{
Values of $\beta$ as defined in 
Eq. (\protect{\ref{definesbeta}}), for $l = 3$,
$3 \leq n \leq 10$,
presented as in Table \protect{\ref{table1}}
}
\label{table3}
\begin{tabular}{rrrr}
\br
$n = 3$&4&5&6\\
\mr
-3.151022588503&-4.406309629179&-5.569879138144&-6.679108747771\\
-0.688273381129&-2.602022794423&-4.214093797042&-5.637934713748\\
3.839295969632&0.834376786655&-1.552972495811&-3.541460362647\\
&6.173955636947&2.678798505483&-0.137921230589\\
&&8.658146925515&4.750180241001\\
&&&11.246244813755\\
\end{tabular}
\begin{tabular}{rrrr}
\br
$n = 7$&8&9&10\\
\mr
-7.754622190689&-8.808385081533&-9.847641443106&-10.876940137334\\
-6.939787368805&-8.160018019327&-9.324116930842&-10.448678615450\\
-5.263146046115&-6.801150749358&-8.209581772798&-9.524662708154\\
-2.490737352918&-4.517404302068&-6.309419903787&-7.929266399940\\
1.551042361196&-1.142156136416&-3.465931796518&-5.514770174927\\
6.987662691921&3.449539586975&0.443287284785&-2.162791993620\\
13.909587905409&9.350359497993&5.511558811618&2.219325513867\\
&16.629215203734&11.809781767444&7.703479769014\\
&&19.392063983204&14.345470691179\\
&&&22.188834055364\\
\end{tabular}
\end{table}

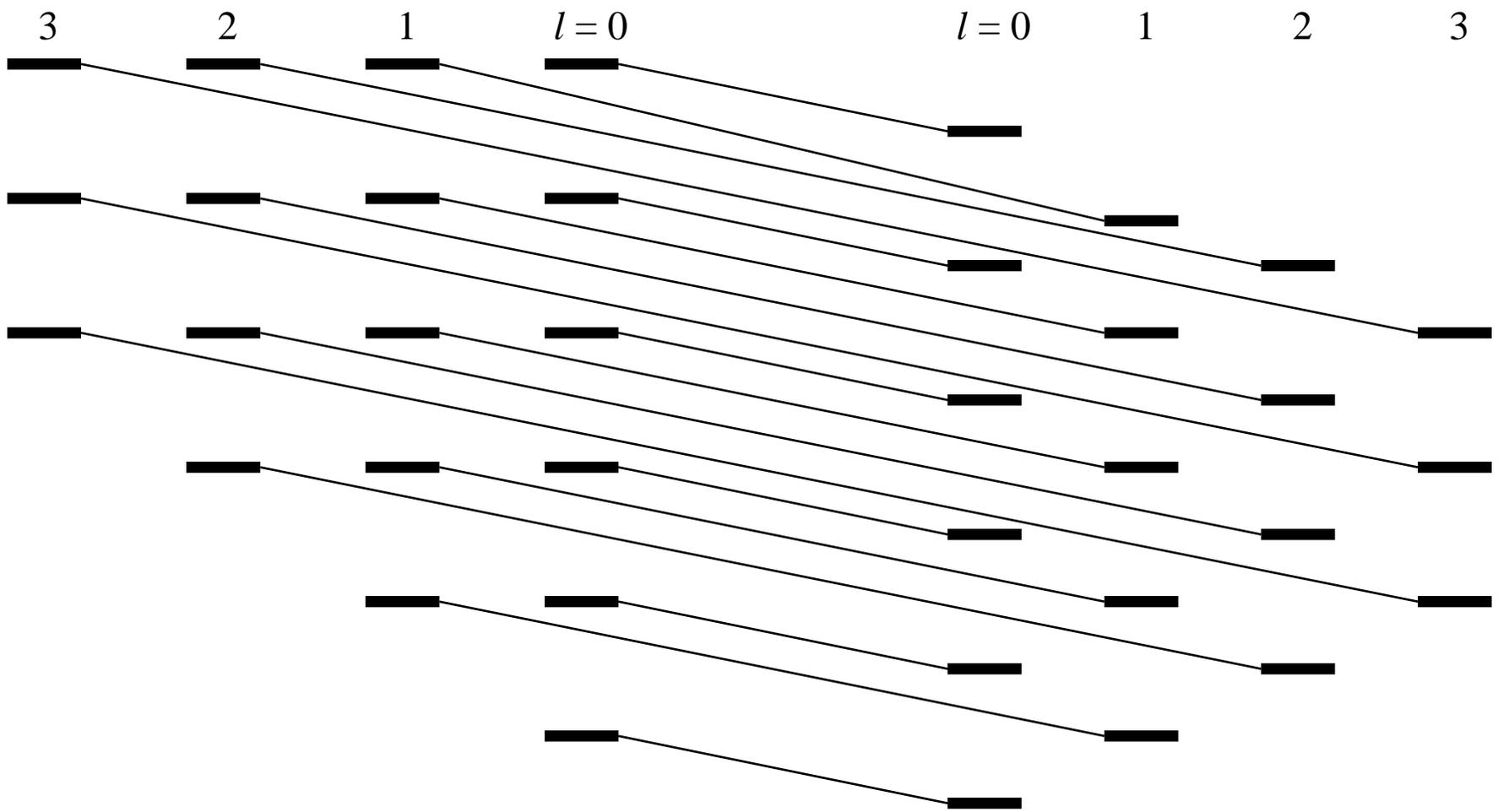
\end{document}